\documentstyle[aps,prl,multicol,epsf]{revtex}
\def\gs{\gtrsim}
\def\ls{\lesssim}
\def\be{\begin{equation}}
\def\en{\end{equation}}                  
\def\p{\partial} 
\newcommand{\bi}[1]{\mbox{\boldmath$#1$}}
\newcommand{\av}[1]{\langle{#1}\rangle}

\def\bea{\begin{equation}\begin{array}{rcl}}
\def\ena{\end{array}\end{equation}}

\def\q{{\footnotesize{\it q}}\kern -5pt {\footnotesize{\it q}}}
\def\k{{\footnotesize{\it k}}\kern -5pt {\footnotesize{\it k}}}
\def\seq{\sim \kern -12pt \lower 5pt \hbox{$\displaystyle =$}}
\def\nnabla{\nabla\kern-3.3mm\nabla}

\def\ge{> \kern -12pt \lower 5pt \hbox{$\displaystyle =$}}
\def\le{< \kern -12pt \lower 5pt \hbox{$\displaystyle =$}}
\def\gs{> \kern -12pt \lower 5pt \hbox{$\displaystyle{\sim}$}}
\def\ls{< \kern -12pt \lower 5pt \hbox{$\displaystyle{\sim}$}}

\def\be{\begin{equation}}
\def\bea{\begin{eqnarray}}
\def\en{\end{equation}}
\def\ena{\end{eqnarray}}
\def\p{\partial }
\def\ve{\varepsilon} 
\renewcommand{\theequation}{\arabic{section}.\arabic{equation}}

\begin{document}
\draft
\bibliographystyle{prsty}
\title{Solvation Effects in Near-Critical Binary Mixtures }
\author{Akira Onuki and Hikaru Kitamura}
\address{Department of Physics, Kyoto University, Kyoto 606-8502, Japan}
\maketitle

\begin{abstract}
{A Ginzburg-Landau theory is presented to investigate 
 solvation effects  in near-critical 
 polar fluid binary mixtures. Concentration-dependence of the 
dielectric constant gives rise to  a  shell 
region around a charged particle 
within which solvation occurs preferentially. 
As the critical point is approached,
 the concentration has  a long-range 
Ornstein-Zernike tail representing strong critical 
electrostriction.   If salt is added, 
strong coupling arises  among the critical fluctuations and 
the ions. The structure factors of the 
critical fluctuations and the charge density 
are calculated and the phase transition behavior is discussed.  
}
\end{abstract}

\begin{multicols}{2}

\section{Introduction}

Solvation effects are of great importance in understanding 
the degree of solubility of ions in various polar solvents 
\cite{Is,Marcus}. A large number of 
papers have been devoted on 
this problem \cite{R1,RK,R2,R3}.  The theoretical 
approaches range 
from solving phenomenological 
continuum models to performing  
computer simulations  on microscopic 
models. Originally,  Born   used a simple continuum theory 
of electrostatics to derive  the solvation (polarization) 
free energy of a single ion with charge $Ze$ 
\cite{Is,Born}, 
\be 
\Delta G_{\rm B}= -({Z^2e^2}/{2R_{\rm B}})
(1- {1}/{\ve}), 
\label{eq:1.1} 
\en 
where   ${R_{\rm B}}$  
represents  the  ionic radius \cite{Is,RMarcus}. 
In his theory the solvent dielectric constant 
$\ve$ is assumed to be homogeneous 
and the contribution without polarization 
($\ve=1$) or in vacuum  is subtracted.  
In the line of continuum 
models,  considerable efforts have  
been made to take into account 
the dielectric saturation near the ion core
(which are particularly crucial for multivalent ions) 
\cite{Abraham,nonlinear}. 
The Born theory also 
neglects  possible inhomogeneity of $\ve$ 
in the vicinity of 
the ion  due to a change 
in the density for one-component fluids 
or in the concentration for binary fluid mixtures 
(electrostriction) \cite{Padova,Wood,Luo,Maryland}.

We point out that the previous  theories have not yet 
treated  the physics 
on  mesoscopic scales 
such as solvation effects on phase transitions 
or collective phenomena at non-dilute salt concentrations. 
In this paper we will show that 
such  effects can be conveniently 
studied  within a scheme of     Ginzburg-Landau 
theory. First, we will apply this approach to 
solvation  around a charged particle 
in  polar binary mixtures near 
the consolute critical point. 
Here important is  the gradient free energy 
well-known in the field of critical 
phenomena \cite{Onukibook}, 
which will  reasonably accounts for   the 
long-range electrostriction around ions. 
In the previous theories 
\cite{Padova,Wood,Luo} in the physical chemistry,   
the gradient free energy has been 
neglected. On the basis of these results we 
will then discuss the effects of  low-density salt  
in polar binary mixtures in a 
Debye-H$\ddot{\rm u}$ckel type approximation. 
This second part is an extension of 
a previous paper by one of the present authors \cite{NATO}. 
We also note  that 
similar  effects have recently 
been discussed when the solvent is 
a liquid crystal \cite{OnukiL}, 
where the  director field can be much deformed 
around a charged particle over a long distance.

\section{Solvation around a single 
charged particle}

\setcounter{equation}{0}

\subsection{Electrostatics}

We  place a charged particle  
in  a near-critical   binary mixture 
in a one-phase state in equilibrium,  
in which the dielectric constant 
$\ve=\ve(\phi)$ depends 
strongly on the concentration. 
This is the case when the two components 
have very different dielectric constants, 
$\ve_A$ and $\ve_{ B}$, with $\ve_A$ considerably larger than 
$\ve_{B}$. There is no established theory of   
the dielectric constant of such mixtures. 
However, empirically, $\ve$ can be  expressed 
roughly as a linear function of the concentration in many 
relevant mixtures \cite{Marcus}. 
In  the present  theory, for simplicity,   we assume 
the linear dependence,
\be 
\ve=\ve_0+\ve_1\phi,  
\label{eq:2.1}
\en 
where  $\ve_0$ and $\ve_1$ are  constants 
(with $\ve_{B}= \ve_0$ and $\ve_A=\ve_0+\ve_1$) 
and $\phi$ is the concentration or the volume fraction  
of the preferred  
component $A$.  Debye and Kleboth \cite{Debye-Kleboth} 
performed a light scattering experiment 
on a mixture of  nitrobenzene (NB)+ 2,2,4-trimethylpentane 
in electric field, for which $\ve(\phi)$ increased   from 
 2.1 to 34.2 with increasing  NB 
with $\p^2\ve/\p \phi^2=28.7$ near the critical point. 
The measured  curve of $\ve(\phi)$ vs $\phi$ may  be 
fairly fitted  to the linear form (2.1) 
(and fitting is 
better in the region $\phi \gs \phi_{\rm c}$).  
In their experiment 	
  the critical temperature was 
29.16 C and the critical 
concentration was 51.7 wt$\%$ of NB 
(corresponding to a volume fraction of 0.381 of NB).
In this paper the dielectric constant is  
  assumed to be 
independent of electric field, although this 
is a questionable assumption 
in the vicinity of small ions 
\cite{Abraham,nonlinear}.

The electric field 
 ${\bi E}=-\nabla\Phi$ is 
  induced by the electric charges and 
the electric potential $\Phi$ 
satisfies 
\be 
\nabla\cdot\ve\nabla \Phi=
 -4\pi \rho({\bi r}),  
\label{eq:2.2}
\en  
where $\rho({\bi r})$ is the charge density. 
In our problem the space-dependence of 
$\ve$ is crucial.

\subsection{Ginzburg-Landau Free Energy}

 We construct a  simple theory for 
 a nearly incompressible 
mixture near the consolute critical point at a given 
pressure. Then $\psi=\phi-\phi_{\rm c}$ is 
the order parameter of 
liquid-liquid phase transition, 
where $\phi_{\rm c}$ is  the critical 
concentration. The total Ginzburg-Landau free energy 
of the system is given by  
\be 
F= \int d{\bi r} 
  \bigg [ f+ \frac{C}{2}|\nabla \phi|^2 
+\frac{\ve}{8\pi} {\bi E}^2 \bigg ] .  
\label{eq:2.3}
\en 
The molecules of the two components  
have a common volume $v_0= a^3$, although 
they can have different volumes in general. 
The free energy density $f=f(\phi)$ is given by 
the Hildebrand \cite{Hil}
(or Bragg-Williams \cite{Onukibook}) 
expression,  
\be 
f= \frac{k_{\rm B}T}{v_0} 
\bigg [ \phi \ln\phi + (1-\phi)\ln (1-\phi) 
+ \chi \phi (1-\phi)\bigg ] , 
\label{eq:2.4}
\en 
where $\chi$ is the  interaction parameter 
dependent on the temperature $T$. 
The critical-point values of 
 $\phi$ and $\chi$ are   
\be 
\phi_{\rm c}=1/2, \quad 
 \chi_{\rm c}=2, 
\label{eq:2.5}
\en 
respectively, in the mean field theory. 
The order parameter is defined by   
\be 
\psi=\phi-1/2.
\label{eq:2.6}
\en 
The coefficient $C$ of the gradient  term 
is of order $k_{\rm B}T/a$ and we 
will use a simple form \cite{Safran,commentG},  
\be 
C= k_{\rm B}T  \chi/a .
\label{eq:2.7}
\en 
The last term in $F$ is the 
electrostatic contribution. 
The free energy functional (2.3) supplemented 
with (2.2) can be used generally 
in the presence of  ions 
in polar near-critical fluids,

\subsection{Preferential solvation }

We  consider a single charged particle at the origin 
of the reference frame. 
The charged particle has a radius $R$. 
It is convenient to  assume that  
the charge density is homogeneous within 
the sphere as 
\be 
\rho = Ze /( 4\pi R^3/3)  \quad (r<R) 
\label{eq:2.8}
\en  
and vanishes outside the sphere $r>R$.  
The total charge is given by $Ze$. 
In our theory $R$ is a 
phenomenological parameter. Then 
the volume fraction $\phi$ and 
the electric potential $\Phi$ can be defined 
even within the sphere 
and, for the single charge 
case,  all the quantities 
may be assumed to depend on space as 
\be 
\phi=\phi(r),\quad  \Phi= \Phi(r),\quad  
{\bi E}= - \Phi'(r) r^{-1}{\bi r},
\label{eq:2.9}
\en
where $r=|{\bi r}|$ and $\Phi'= d\Phi/dr$.   
The concentration $\phi(r)$ 
decrease from $\phi(0)$ to $\phi_\infty$ 
with increasing $r$ (see Fig.1 to be presented below). 
In this case (2.2) is solved to give 
\be 
-\ve (\phi) \Phi' (r) 
= \frac{4\pi}{r^2} \int_0^r   dr_1 r_1^2 \rho(r_1)
= \frac{Ze}{r^2}\theta(r),  
\label{eq:2.10}
\en 
where $\theta(r)=1$ for $r>R$ and  
\be 
\theta(r)= (r/R)^3\quad (r<R)
\label{eq:2.11}
\en  
under the assumption (2.8). 
The electrostatic free energy  
is now simply  of the form, 
\be 
\int d{\bi r} \frac{1}{8\pi}\ve {\bi E}^2
= \frac{1}{2} Z^2e^2 
\int_0^\infty  d{r} \frac{1}{\ve r^2}\theta(r)^2 
\label{eq:2.12}
\en

If $\ve$ is  replaced by 
the dielectric constant  far from 
the ion $\ve_\infty$, 
the above quantity is integrated to 
become a Born contribution (see (1.1))\cite{Born}, 
\be 
F_{{\rm B}\infty}= Z^2e^2/2{\ve}_\infty 
R_{\rm B}. 
\label{eq:2.13} 
\en 
In terms of the critical-point 
dielectric constant  $\ve_{\rm c}=\ve_0+\ve_1/2$ 
 and the order parameter
 $\psi_\infty=\phi_\infty-1/2$ far from the ion, 
we have 
\be 
\ve_\infty= \ve_0+\ve_1\phi_\infty=
\ve_{\rm c}+ \ve_1\psi_\infty.  
\label{eq:2.14}
\en 
The Born radius is given by 
\be 
R_{\rm B}= 5R/6.
\label{eq:2.15} 
\en 
In the integral (2.12) the contribution 
from  the ion interior  $r<R$ 
is $1/5$ of that from  the ion exterior, resulting in 
 the factor $1/(1+1/5)= 5/6$ in (2.15). 
However, if we replace $\ve$ by its value at the ion 
$\ve(0)=\ve_0+\ve_1\phi(0)$, we obtain another estimate, 
\bea 
F_{{\rm B}0}&=& Z^2e^2/2\ve(0)R_{\rm B}\nonumber\\
&=& 
F_{{\rm B}\infty}\ve_c/[\ve_0+\ve_1\phi(0)]. 
\label{eq:2.16} 
\ena 
As will be calculated in the appendix,
 the  equilibrium free energy of 
a single ion accounting for  the inhomogeneous 
$\phi$ and $\ve$ in the mean field theory reads   
\bea 
F_{\rm sol}&=& 4\pi \int_0^\infty 
 dr r^2 [ \hat{f}+ C(\phi')^2 /2+\ve {\bi E}^2/8\pi ]
\nonumber\\
&=& \frac{4\pi}{3} 
\int_0^\infty  dr  \bigg 
[Cr^2 (\phi')^2 + \frac{Z^2e^2\theta^2}{8\pi r^4} 
 \frac{d}{dr} \bigg (\frac{r^3}{\ve }\bigg ) \bigg ], 
\label{eq:2.17}
\ena 
where 
\be 
\hat{f}(\phi)= f(\phi)- f(\phi_\infty)- 
\mu (\phi-\phi_\infty), 
\label{eq:2.18} 
\en 
with $\mu=  f'(\phi_\infty)$ being the chemical 
potential difference.  
As shown in the second line of (2.17), 
$F_{\rm sol}$  consists of 
the gradient contribution ($\propto C$) 
and the electrostatic contribution ($\propto e^2)$. 
The latter becomes  the Born contribution 
(2.13) or (2.16) if  $\ve$ is replaced by  
$\ve_\infty$ or $\ve(0)$.  In Fig.7 below 
we shall see the relation 
$F_{{\rm B}0}<F_{\rm sol}<F_{{\rm B}\infty}$.

With the aid of  (2.11) 
the minimization condition 
$\delta F/\delta \phi=\mu=$const.  in equilibrium is 
rewritten as 
\be
\ln \bigg [\frac{1+ 2\psi}{1-2\psi}\bigg ]  - 2\chi \psi - 
\chi a^2\nabla^2 \psi 
-\nu 
=  A\frac{a^4 \theta^2}{{\hat{\ve}}^2 r^4}   ,
\label{eq:2.19}
\en 
where  $\hat{\ve}$ is the normalized dielectric constant, 
\be 
\hat{\ve}= \ve/\ve_{\rm c}= 1+ (\ve_1/\ve_{\rm c}) \psi.  
\label{eq:2.20}
\en 
The dimensionless chemical potential difference 
${\nu}= v_0 \mu/k_{\rm B}T $ is written in terms of 
$\psi_\infty=\psi(\infty)$ as 
\be 
{\nu}=  \ln[(1+2\psi_\infty)/(1-2\psi_\infty)]
 - 2\chi \psi_\infty.
\label{eq:2.21}
\en 
The coefficient $A$ on the right-hand side 
of (2.19) represents the strength of solvation   
 defined by 
\bea 
A&=& \ve_1 Z^2e^2/8\pi \ve_{\rm c}^2 k_{\rm B}Ta \nonumber\\
&=&\ve_1 Z^2\ell_{\rm B}/8\pi \ve_{\rm c} a  .  
\label{eq:2.22}
\ena 
In the second line the Bjerrum length  
\be 
\ell_{\rm B}= e^2/\ve_{\rm c}k_{\rm B}T
\label{eq:2.23} 
\en 
at the critical concentration is introduced.  
For example, if we set $\ell_{\rm B} \sim 14\AA$, $a\sim 2 \AA$, 
$\ve_1/\ve_0\sim 2$, we obtain $A \sim Z^2/5$. 
In the vicinity of the ion 
the right-hand side of (2.19) takes a  maximum 
of order $Aa^4/R^4$ (because $\hat{\ve} \sim 1$). 
If this maximum is much larger than 1, 
there arises a solvation shell around the ion  
within which preferential solvation is strong 
and $\phi \cong 1$. Thus the condition of strong solvation 
may be expressed as 
\be 
A> (R/a)^4.
\label{eq:2.24} 
\en 
The solvation is stronger for smaller $R/a$, which is a well-known 
result in the literature \cite{Is}.

It is worth noting that 
Padova presented an equation similar to (2.19) for 
the concentration in the vicinity of an 
ion using thermodynamic 
arguments  \cite{Padova}, 
where  $C=$ asnd the gradient term ($\propto \nabla^2\psi$) 
did  appear. Even far from the critical point, 
as can be  seen  in Fig.8 below, 
 the result with  $C=0$  
is considerably  different from that with   
the gradient term 
in the strong solvation case.

\subsection{Relations near the critical point}

If the system is near  the critical point far from 
the charged particle, we have $\chi \cong  2$ and 
$\phi_\infty \cong 1/2$. Then the inverse 
correlation length $\kappa=\xi^{-1}$ far from the ion 
is determined  by 
\bea 
a^2\kappa^{2} &=&  
{4}/{(1-4\psi_\infty^2)\chi}-2 \nonumber\\
&\cong& 
2-\chi +8 \psi_\infty^2.
\label{eq:2.25}
\ena    
As $T \rightarrow T_{\rm c}$ in the UCST case we may set 
\be 
2-\chi \cong D_1(T/T_{\rm c}-1).
\label{eq:2.26} 
\en   
The correlation length in  
 the mean-field expression becomes  
$
\xi=\xi_{0+}(T/T_{\rm c}-1)^{-1/2}$ 
with 
\be 
\xi_{0+}= aD_1^{-1/2}
\label{eq:2.27} 
\en 
in one-phase states at the critical 
composition \cite{Debye-Kleboth}.  
The fluid can remain  near-critical 
only in the region where 
$Aa^4/r^4 \ll 1$ or 
\be 
r\gg A^{1/4}a.
\label{eq:2.28} 
\en 
The shell radius should thus grow 
as $A^{1/4}a$ for $A \gg 1$. 
In the near-critical region,     
(2.19) may be approximated in the 
 Landau expansion form  \cite{NATO}, 
\be 
a^2[\kappa^2 -\nabla^2]\delta\psi 
+ \frac{8}{3}(3\psi_\infty\delta\psi^2 +\delta\psi^3) 
= A\frac{a^4\theta^2}{2 r^4} , 
\label{eq:2.29}
\en
where $\delta\psi=\psi-\psi_\infty$. 
This equation does not hold within the solvation shell 
if $\psi$ is not small there for strong solvation, but 
 holds generally  far from the ion. 
If we retain the linear terms only in the above  equation, 
we obtain  $(\kappa^2-\nabla^2)\delta\psi \cong 0$ 
far from the ion,   so 
 there should  emerge the Ornstein-Zernike tail, 
\be 
\delta\psi(r) \cong  \frac{B}{r}e^{-\kappa r} , 
\label{eq:2.30}
\en 
in the concentration. The coefficient $B$ is a microscopic 
length to  be determined 
in the following.

In the  Landau theory of 
phase transition  (usually presented 
in the context of Ising spin  
systems) \cite{Onukibook},  
the right-hand side of (2.19) or (2.29) 
plays the role of 
a (position-dependent) magnetic field $h$. 
If $h$ were homogeneous and small, we would have 
the linear response $\delta\psi\sim  h/\kappa^2$. 
In the present case, however, 
$h=h(r)$  steeply changes in space 
($\propto r^{-4}$)  from 
a maximum of order $Aa^4/R^4$ to zero 
and as a result the  gradient term 
$-\nabla^2\psi$ becomes increasingly  crucial 
with decreasing $\kappa$.

\subsection{Weak solvation limit near the critical point}

It is instructive first to  consider 
the simplest case of very small $A$  
 near the critical point
where   $0<2-\chi\ll 1$  
 and $\psi_\infty=0$.  
Furthermore, if $\psi(0) \ll \kappa a$ holds at the ion, 
we may neglect the second  
 term on the left-hand side of   (2.29) 
 in the whole space 
to obtain the linearized equation, 
\be 
(\kappa^{2}-\nabla^2)  \psi= 
A {\theta(r)^2a^2}/{2r^4} , 
\label{eq:2.31}
\en 
where we  have set $\hat{\ve}=1$. 
This equation is readily integrated to give 
\be 
 \psi(r)= \frac{Aa^2}{4\kappa r} 
 \int_0^\infty \frac{dr_1}{r_1^3} 
 e^{-\kappa|r-r_1|}{\theta(r_1)^2} 
- \frac{D}{r}e^{-\kappa r}, 
\label{eq:2.32}
\en 
where the coefficient $D$ is determined such that 
$\psi(0)$ is finite. We can see the relations,  
\bea 
\psi(0)&=& 2\kappa D 
= {A} a^2 \int_0^\infty {dr_1} 
 e^{-\kappa r_1} {\theta(r_1)^2}/2r_1^3 \nonumber\\
&\cong&3Aa^2/8R^2 \quad (\kappa R \ll 1).
\label{eq:2.33}
\ena 
The  second line holds for $\kappa R \ll 1$. 
As stated above (2.31), the linearization  is allowed for 
\be 
A \ll \kappa R^2/a \quad {\rm or}\quad  
\ve_1 Z^2\ell_{\rm B}/8\pi \ve_{\rm c}\ll R^2 \kappa  .  
\label{eq:2.34}
\en

From (2.31) the three-dimensional 
Fourier transformation of 
$\psi$ becomes $\psi_{\small{\bi k}}= 
H_k/ (\kappa^2+k^2)$ where 
\be 
H_k= 2\pi Aa^2\int_0^\infty  d{r}\theta(r)^2\sin(kr)/kr^3. 
\label{eq:2.35}
\en 
Here $H_k$ tends to  a well-defined  long-wavelength limit 
$H_0=\lim_{k\rightarrow 0}H_k$  for $kR\ll 1$.  
Thus,  for $\kappa  R\ll 1$, 
 $\psi(r)$ approximately  takes  the Ornstein-Zernike 
form (2.30) in a wide  space  region with 
  $B=a^2H_0/4\pi$. Here,  
\be 
B=  Aa^2\int_0^\infty dr \theta(r)^2/2r^2=Aa^2/2R_{\rm B}, 
\label{eq:2.36}
\en  
with $R_{\rm B}$ being given by (2.15).  
This form also directly 
follows from the integral form 
(2.32) (by taking 
the limit $\kappa \rightarrow 0$).  
If we do not assume  the special form  $C$ 
 in (2.7),  we have 
$B= (k_{\rm B}T/C)Aa/R_{\rm B}$ 
(which will be used in (3.12) below).

At very long distances $r>\lambda$, 
however, the gradient term $-\nabla^2\psi$ 
in (2.31) becomes negligible to give  
\be 
\psi(r) \cong {Aa^2}/{2\kappa^2 r^4} .
\label{eq:2.37}
\en 
This is   the relation of  linear response 
as stated at the end of subsection 3D. 
The crossover length $\lambda$ 
is longer than $\xi=\kappa^{-1}$ 
and is determined by the 
 balance of the Ornstein-Zernike form (2.30) 
and the long-distance limit (2.37) as 
\be 
(\kappa\lambda)^3 e^{-\kappa \lambda} = \kappa R .
\label{eq:2.38}
\en  
For example, $\lambda \cong 10.8 \xi$ for 
$R=0.03\xi$. 
Thus $\kappa\lambda$ increases as 
$\kappa R$ decreases.

We may then calculate the excess accumulation $\Psi$ 
 of the component A around the ion as 
\bea 
\Psi&=& \int d{\bi r} \psi(r)\nonumber\\
&=& 2\pi Aa^2 /\kappa^2 R_{\rm B}= 4\pi B/\kappa^2 ,
\label{eq:2.39}
\ena 
which coincides with 
the space integral of the Ornstein-Zernike form  (2.30) 
and grows  strongly on approaching the critical point.
The integral of the tail (2.38) in the region $r>\lambda$ 
is much  smaller than the value in (2.39) 
by $R/\lambda$ and is canceled 
with a contribution from the correction 
to the Ornstein-Zernike form in the range $r<\lambda$.

\subsection{Behavior at the critical point}

At the critical point 
($\chi=2$ and $\phi_\infty=0.5$),  
$\psi(r)$ is  of the form $Ba/r$ at long distances 
in the linear theory. We here show that  it is 
delicately modified  in the nonlinear theory. 
We rewrite (2.29)  for $r\gg a$ as 
\be 
-a^2\nabla^2\psi 
+ \frac{8}{3}\psi^3 
= A\frac{a^4}{2 r^4} .
\label{eq:2.40}
\en
We set $G(s)= r\psi(r)/a$ with 
$s=\ln(1+r/a)$ to obtain 
\be 
(1-e^{-s})^2 \bigg 
[\frac{d^2}{ds^2}-\frac{d}{ds}\bigg ] G(s)
=  \frac{8}{3}G(s)^3 -\frac{1}{2}Ae^{-s}  .
\label{eq:2.41} 
\en 
For $s\gg 1$ we assume  the algebraic decay  
$G(s) \sim s^{-\beta}$. Then 
the above equation is approximated as 
$-d G(s)/ds \cong 8 G(s)^3/3$, yielding  
the solution
\be  
G(s) \cong (3^{1/2}/4)s^{-1/2}
\label{eq:2.42} 
\en 
with $\beta=1/2$.  
Thus the asymptotic behavior at extremely long distances 
$s \gg 1$ should be 
\be 
\psi (r) \cong  {3^{1/2}a}/[4r \sqrt{\ln(1+r/a})] . 
\label{eq:2.43} 
\en 
The above form 
will turn out  to be 
consistent with numerical analysis in Fig.6 
to be presented below.

\subsection{Numerical Results}

We then show numerical solutions of 
(2.19) around an ion in 
equilibrium one-phase states for the case  
$\ve_0=10$ and $\ve_1=70$ in (2.1), 
assuming  the model charge distribution (2.8). 
The ion radius is given by $R=0.3a$ in 
terms of the solvent radius $a$.

We show the concentration profile,  
$\phi(r)$ vs $r/a$,  in Fig.1 
and the concentration $\phi(0)$ 
at the ion center  in Fig.2. Notice that $\phi(r) \cong \phi(0)$ 
within the  interior of the 
ion $r \ls R$  owing  to the smoothing effect 
of the gradient term. 
We can see that the concentration $\phi(0)$ at  the 
ion increases  obeying 
 the linear growth  (2.31) for $A\ll 1$ and 
 saturates into  1 for $A \gs 0.3$  
and that the shell region expands 
with increasing $A$.  The shell radius $R_{\rm shell}$ 
may be defined such that  
$\phi(r)>0.75$ for $r<R_{\rm shell}$, say. 
Then $R_{\rm shell} \sim A^{\alpha}$ 
with the effective exponent 
$\alpha$ being about $1/4$ for 
the profiles (d)$\sim$(g) in Fig.1. Below  (2.28)   
we have discussed that   
$\alpha$ should tend to $1/4$ for $A \gg 1$.

In Fig.3 we examine the behavior of 
$\psi(r)= \phi(r)-1/2$ with $\phi_\infty=1/2$ at $A=0.346$. 
We can see that the combination  $r\psi(r)/a$  
behaves as $(B/a) \exp(-\kappa r)$ 
in a wide intermediate range $a \ls r\ls \lambda$. 
The correlation length  $\xi=\kappa^{-1}$ is 
given by $10a$ for $\chi=1.99$ and 
by $10^{3/2}a$  for $\chi=1.999$.  For $r\gs \lambda$, $\psi(r)$ 
can be fitted to the $r^{-4}$ tail in (2.37).  
Fig.4  displays  $B/a$ vs $A$ 
for $\chi=1.99$ and $1.999$ at $\phi_\infty=0.5$. 
In the nonlinear regime $A\gs 0.1$ 
the data may be fitted to $B/a \sim 
A^{0.4}$. 
In Fig.5 we plot the crossover length $\lambda$ divided by $a$ 
obtained from equating the two fitted 
functions (2.30) and (2.37). In the linear regime with 
small $A$, $\lambda$ is given by (2.37).

The concentration profile just at the critical point 
is also of great interest. In Fig.6 we plot the combination 
$\psi(r)rs^{1/2}$ vs $s=\ln(1+r/a)$ in the region $s<12$ 
or $r/a<e^{12}-1$ for $A=0.2,1,$ and 5. 
If (2.42) is valid, this combination  should tend 
to $3^{1/2}/4 \cong 0.43$.  Thus the three curves are 
consistent with (2.42).

 Fig.7 displays  the numerically 
calculated solvation free energy $F_{\rm sol}$ in (2.17) 
divided by $F_{\rm{B}\infty}$ in (2.13). 
Also plotted is the ratio 
$F_{\rm{B}0}/F_{\rm{B}\infty}= \ve_c/[\ve_0+\ve_1\phi(0)]$ in 
(2.26).  Thus we find that $F_{\rm sol}$ is intermediate between 
$F_{\rm{B}0}$ and $F_{\rm{B}\infty}$. 
Coincidence of the two curves 
of $\chi=1.99$ and $1.999$ 
implies  that $F_{\rm sol}$ tends 
to a constant on approaching  the critical 
point.

In Fig.8
we demonstrate that 
$r\psi(r)/a$ vs $r/a$ 
 far above  the critical point 
with  $\chi=1.5$ and $\phi_\infty=0.5$.
The profile can be fitted neither to
 the Ornstein-Zernike 
form  nor to 
the solution of (2.19) 
without the gradient  term. 
The curves  from the local equilibrium theory 
without the gradient  term 
\cite{Padova} considerably differs 
 from that in our theory  even away from 
the critical point.

\setcounter{equation}{0}
\section{Near-critical fluids with ions}

Experimentally,
 it has long been  known that  even a small 
fraction  of ions (salt) 
 with $c \ll 1$  dramatically changes  the liquid-liquid 
phase behavior, where  $c$ is 
the mass or mole  fraction of ions. 
For small $c$, the UCST coexistence curve 
 shifts upward as 
\be 
\Delta T_{\rm ion } = A_{\rm ion} c + O(c^2)
\label{eq:3.1}
\en 
with large positive coefficient $A_{\rm ion}$, expanding   
 the region of demixing. 
For example, $A_{\rm ion}/T_{\rm c} \sim 10$ 
with $T_{\rm c} \sim 300$K   when  
  NaCl  was added to  
cyclohexane$+$methyl alcohol \cite{polar1} 
and to triethylamine+H$_2$O \cite{polar2}.  
The LCST coexistence curve of  2,6-lutidine+H$_2$O(D$_2$O), 
the shift is downward with  $|A_{\rm ion}|/T_{\rm c} \sim 10$ 
\cite{newly_found}. 
Similar large impurity effects 
were observed when water was added to 
methanol+cyclohexane \cite{polar3}. 
In some aqueous  mixtures, 
even  if they are  miscible at all $T$ 
at atmosphere pressure without salt, addition of  
a  small amount of  salt 
gives  rise to reentrant phase separation behavior 
 \cite{Kumar,cluster,Anisimov,Misawa}. Such reentrant 
phase behavior is believed to arise from 
hydrogen bonding.

\subsection{Ginzburg-Landau free energy}

In the previous section we have examined 
the concentration profile 
around a single fixed charged particle. 
We then need to construct a theory 
on polar binary mixtures in which 
a small amount of salt is doped \cite{NATO}. 
Even if the ion concentration is very 
small, the interactions among the ions 
should be much influenced by 
 the preferential solvation near the critical point. 
We here  present a Ginzburg-Landau free energy 
in which the  fluctuations 
on the spatial scale of the solvation shell 
radius  $R_{\rm shell}$ have been coarse-grained. 
Then the order parameter 
$\psi({\bi r})=\phi({\bi r})-\phi_{\rm c}$ 
and the ion densities 
$n_1({\bi r})$ and $n_2({\bi r})$ 
for the two species do not 
change appreciably on the scale of $R_{\rm shell}$.

The  two ion species, 1 and 2,
 have  charges, 
$Q_1=Ze$ and $Q_2=-e$. 
The average densities are 
written as 
\be 
\av{n_1}={\bar n}, \quad 
\av{n_2}=Z{\bar n}, 
\label{eq:3.2}
\en 
where $\av{\cdots}$ denotes 
taking the  space average. 
Notice that the charge neutrality 
condition yields $Z \av{n_1}= \av{n_2}$. 
The total charge number density is written as 
\be 
n_{\rm tot}= (Z+1) {\bar n}.
\label{eq:3.3}
\en 
We propose the  free energy 
$F$  in the form \cite{NATO}, 
\bea 
F&=& \int d{\bi r} 
  \bigg [ f(\psi) + \frac{C}{2}|\nabla \psi|^2 
+\frac{\ve_{\rm c}}{8\pi} {\bi E}^2  \nonumber\\ 
&&+k_{\rm B}T 
  \sum_{K=1,2}( 
n_K \ln n_K +  w_Kn_K\psi) \bigg ] .
\label{eq:3.4}
\ena 
The first line is of the same form as 
the free energy in (2.3), but $f(\phi)$ can 
be expressed in the Landau expansion form 
$f=r\psi^2/2+ u_0\psi^4/4$ because the non-critical 
behavior in the solvation shell is not treated here. 
In addition  we do not assume the special form (2.7) for 
the coefficient $C$. The 
electric field ${\bi E}= -\nabla\Phi$ here arises 
from the Coulomb interaction among the ions so that 
\be 
\nabla\cdot{\bi E}= -\nabla^2\Phi= 4\pi \ve_{\rm c}^{-1} 
 \rho({\bi r}) ,
\label{eq:3.5}
\en 
where 
\be 
\rho({\bi r})= e[Zn_1({\bi r})-n_2({\bi r})] 
\label{eq:3.6}
\en 
is the charge density. Therefore, the 
electrostatic part of the free energy  
is rewritten as 
\be 
\int d{\bi r}\frac{\ve_{\rm c}}{8\pi} {\bi E}^2= \frac{1}{2\ve_{\rm c}}
\int d{\bi r}\int d{\bi r}' 
\rho({\bi r})\rho({\bi r}')\frac{1}{|{\bi r}-{\bi r}'|} .
\label{eq:3.7}
\en 
Because the small-scale electrostriction 
on the scale of $R_{\rm shell}$  
has been  coarse-grained, 
we neglect the small-scale inhomogeneity 
of the dielectric constant and set $\ve =\ve_{\rm c}$ 
with $\ve_{\rm c}$ being the dielectric constant at the 
critical point.

The free energy  with the bilinear coupling  
$(\propto n_K\psi$)
was already examined 
in Refs.\cite{metal,Odessa}, 
 where   the coupling constants $w_K$ were
 phenomenological parameters, however. 
We claim that     
the small-scale solvation 
or electrostriction  is the origin of 
the bilinear coupling. 
 In fact, the minimization of $F$ at fixed 
inhomogeneous ion densities yields 
\be 
\mu= f'-C\nabla^2\psi + k_{\rm B}T 
 (w_1n_1+ w_2n_2)= {\rm const.} ,
\label{eq:3.8}
\en 
where $f'=\p f/\p \psi$. 
For  small inhomogeneous deviations 
$\delta\psi=\psi-\av{\psi}$ and  
$\delta n_K= n_K-\av{n_K}$ ($K=1,2$) we linearize
 the above equation to obtain 
\be 
C( \kappa^2 - \nabla^2)\delta \psi + k_{\rm B}T 
 (w_1\delta n_1+ w_2\delta n_2)= 0, 
\label{eq:3.9}
\en 
where $C\kappa^2= \p^2f/\p\psi^2$ with 
$\kappa$ being the inverse correlation length.  
Notice that, if the ion densities are localized 
in regions shorter than $\xi=\kappa^{-1}$,  
$\delta\psi$ has   Ornstein-Zernike tails 
in the form of (2.30) around such localized regions. 
Thus   $w_K$ is written as  
\be 
w_K= -(4\pi C /k_{\rm B}T)B_K   \quad (K=1,2),  
\label{eq:3.10}
\en 
where  $B_K$ is the coefficient  
of the Ornstein-Zernike tail around an ion of 
the species $K$. If the two components 
of the fluid mixture have the same molecular size $a$ 
as assumed in the previous section, 
 $B_K$ is  determined by 
the parameters $A_K$ in (2.22), where 
\be 
A_1/Z^2=A_2= \ve_1 \ell_{\rm B}/8\pi \ve_{\rm c} a  .  
\label{eq:3.11}
\en  
In particular, 
in the linear solvation regime with small $A_K$, (2.36) yields
\be   
w_K = -4\pi A_K a/R_{{\rm B}K}\quad ({\rm linear}),
\label{eq:3.12}
\en 
where $R_{{\rm B}K}$ is the Born radius of the ion 
species $K$.  See  Fig.4 for the nonlinear behavior of $B_K$. 
That is, $B_K/a  \sim A_K^{0.4}$ for large $A_K$ 
and   $w_K$ can well exceed 1.

In Ref.\cite{NATO} $w_K$ 
 was  related to the Born solvation free energy 
$\Delta G_{{\rm B}K}= -({Q_K^2}/{2R_{{\rm B}K}})
(1- {1}/{\ve})$ in (1.1) which should be 
defined for the two species. 
That is, treating $\ve$  as a function of $\psi$, 
we obtain   
\be 
k_{\rm B}T w_K=  
\frac{\p}{\p \psi}\Delta G_{{\rm B}K}=   
-\ve_1 Q_K^2 /2\ve_{\rm c}^2 R_{{\rm B}K},
\label{eq:3.13}
\en   
where the differentiation 
is taken at $\phi=\phi_{\rm c}$. Obviously, 
(3.12) and (3.13) are equivalent from (2.22). 
This simple derivation cannot be used 
in the strong solvation regime, however.

\subsection{Fluctuations  in one-phase states}

We consider   small  plane-wave fluctuations 
in  a  one-phase state. 
The  fluctuation contributions  to $F$ 
in the  bilinear order  
are  written as 
\bea 
{\delta F}  &=& 
\int_{\bi q} \bigg [
\frac{1}{2}({r+Cq^2})  |\psi_{\small{\bi q}}|^2+
\frac{2\pi }{\ve_{\rm c} q^2}|\rho_{\small{\bi q}}|^2
 \nonumber\\
&&+ k_{\rm B}T \sum_{K=1,2} \bigg (
\frac{|n_{K{\small{\bi q}}}|^2}{2\av{n_K}}  + 
w_K n_{K {\small{\bi q}}}\psi_{\small{\bi q}}^*\bigg )   
\bigg ]  ,
\label{eq:3.14}
\ena 
where $\int_{\bi q}\cdots =(2\pi)^{-3}\int d{\bi q}\cdots$ denotes 
the integration over the wave vector $\bi q$  and    
\be 
r= \p^2f/\p\psi^2.
\label{eq:3.15}
\en 
The  $\psi_{\small{\bi q}}$, 
 $n_{K{\small{\bi q}}}$, and  
$\rho_{\small{\bi q}}$ are  the Fourier 
transformations of $\delta\psi$, 
 $\delta n_K$,  
and  $\rho$, respectively.
We may set 
\be 
r= a_0 (T-T_{\rm c}) 
\label{eq:3.16}
\en 
at the critical concentration, where 
 the coefficient $a_0$ is 
of order $k_{\rm B}/a^3$ and is equal to $4k_{\rm B}D_1/a^3$ 
with $D_1$ being defined by (2.26). 
In the mean-field expression 
$\xi=\xi_{0+}(T/T_{\rm c}-1)^{-1/2}$ 
of the correlation length,  we have 
\be 
a_0 = C  /
\xi_{0+}^{2}T_{\rm c} .
\label{eq:3.17}
\en

We may then calculate the 
structure factor  for the order parameter 
$S(q)= \av{|\psi_{\small{\bi q}}|^2}$ 
in the mean field theory. Since the equilibrium 
distribution is given by const.
$\exp(-\delta F/K_{\rm B}T)$, 
the inverse of $S(q)$ 
 is written as \cite{metal,NATO}   
\bea 
k_{\rm B}T/S(q)&=& a_0(T- T_{\rm c}- \Delta T_{\rm ion}) \nonumber\\
&&+ C q^2 [1- \gamma_{\rm p}^2/(1+\lambda_{\rm D}^2q^2)] , 
\label{eq:3.18}
\ena 
where  $\Delta T_{\rm ion}$ is the critical 
temperature shift due to salt  given by  
\be 
 \Delta T_{\rm ion}
= (k_{\rm B}T_{\rm c}/a_0)(w_1+Zw_2)^2{n}_{\rm tot}/(1+Z)^2 , 
\label{eq:3.19}
\en 
and $\lambda_{\rm D}$ is the Debye screening 
length defined by 
\be 
\lambda_{\rm D}^{-1} = 
(4\pi Z {n}_{\rm tot} e^2/\ve_{\rm c} k_{\rm B}T)^{1/2}.
\label{eq:3.20}
\en 
We introduce a dimensionless  parameter,  
\bea 
\gamma_{\rm p}&=&  |w_1-w_2|(k_{\rm B}T /4\pi C\ell_{\rm B})^{1/2} 
/(1+Z)
\nonumber\\
&=&  {|B_1-B_2|}(4\pi C/\ell_{\rm B}k_{\rm B}T)^{1/2}/({1+Z} ). 
\label{eq:3.21}
\ena 
This number is independent of the ion density 
and represents the strength  of 
asymmetry in  the ion-induced polarization between   the 
two components. 
In the linear solvation regime with small $A_K$,  
we use (3.12) to obtain 
\be 
\gamma_{\rm p} = \frac{\ve_1
\sqrt{\ell_{\rm B} k_{\rm B}T}}{2\ve_{\rm c}(1+Z)\sqrt{4\pi C}} \bigg |
 \frac{Z^2}{R_{{\rm B}1}}-
\frac{1}{R_{{\rm B}2}}\bigg | 
\quad ({\rm linear}).
\label{eq:3.22}
\en
The structure factor of the charge density 
fluctuations $S_{\rho\rho}(q)= \av{|\rho_{\small{\bi q}}|^2}/e^2$ 
is written as 
\bea 
&&S_{\rho\rho}(q)=
 Zn_{\rm tot} \frac{\lambda_{\rm D}^2q^2}{1+\lambda_{\rm D}^2q^2} 
\nonumber\\
&&+ (w_1-w_2)^2\bigg  (\frac{Zn_{\rm tot}}{Z+1}\bigg )^2\bigg (
 \frac{\lambda_{\rm D}^2q^2}{1+\lambda_{\rm D}^2q^2} \bigg )^2 
S(q) ,
\label{eq:3.23}
\ena 
where the first term is the Debye-H$\ddot{\rm u}$ckel 
structure factor 
and the second term arises from the coupling 
to the order parameter.

From  the structure factor $S(q)$ in (3.18) 
we may draw the following conclusions.\\
(i) If $\gamma_{\rm p}<1$, 
$S(q)$ is maximum at $q=0$ and 
the  critical temperature shift due to ions 
is given by (3.19).   As a rough estimate for the monovalent 
case $Z=1$,  we set 
$a_0^{-1} \bar{n} \sim 
k_{\rm B}T \xi_{+0}^3 \bar{n}  \sim {c}$, where 
$ c$ is the mass or mole  fraction of the ions. 
Then  $\Delta T_{\rm ion}
\sim  (w_1+w_2)^2 {c}$. If $|w_1+w_2| \sim 3$, this result 
is consistent with the experiments 
\cite{polar1,polar2}. In future 
experiments,  let $k_{\rm B}T /C S(q)$ vs $q^2$ be
 plotted; then,  
the slope is $1-\gamma_{\rm p}^2$ 
for $q\lambda_{\rm D} \ll 1$ 
and is $1$ for $q\lambda_{\rm D} \gg 1$. 
This changeover should be 
  detectable unless 
$\gamma_{\rm p}\ll 1$.\\
(ii) The case $\gamma_{\rm p}=1$ 
corresponds to a Lifshitz point 
\cite{Lu,Cha}, where 
$1/S(q)-1/S(0) \propto q^4/(1+q^2\lambda_{\rm D}^2)$.\\ 
(iii) If $\gamma_{\rm p}>1$, 
the structure factor 
attains a maximum at an 
intermediate wave number $q_{\rm m}$ given by  
\be 
q_{\rm m}= ( \gamma_{\rm p}-1)^{1/2}/
 \lambda_{\rm D}  ,
\label{eq:3.24}
\en 
so $q_{\rm m} \propto {n_{\rm tot}}^{1/2}$. 
The maximum of the structure factor  
 $S(q_{\rm m})$ is written as 
\be 
S(q_{\rm m})=k_{\rm B}T /[a_0(T- T_{\rm c}- \Delta T_{\rm ion}')] , 
\label{eq:3.25}
\en 
 where 
\be 
\Delta T_{\rm ion}'= \Delta T_{\rm ion} +  
(\gamma_{\rm p}-1)^2T_{\rm c} \xi_{0+}^2 /\lambda_{\rm D}^2 ,
\label{eq:3.26}
\en 
where use has been made of (3.17).  
A  charge-density-wave 
phase should be realized for $T-T_c<\Delta T_{\rm ion}'$.
It is remarkable that this mesoscopic 
phase appears 
however small $n_{\rm tot}$ is 
as long as $\gamma_{\rm p}>1$ and $q_{\rm m} L \gg 1$ with 
$L$ being the system length.  
Here relevant is the coupling of the order parameter 
and the charge density in the form 
$\propto \psi \rho$ in the free energy density, which generally 
exists in ionic systems. 
This    possibility of 
a mesoscopic phase  was first  predicted  
for  electrolytes  by 
Nabutovskii {\it et al}\cite{metal}, 
but has not yet been confirmed in experiments. 
In  polyelectrolytes, on the other hand, 
electric  charges are attached to polymers 
and the structure 
factor of the polymer is known to 
take a  form similar  to 
(3.18), leading to a mesoscopic phase at 
low temperatures, 
 in the  Debye-H$\ddot{\rm u}$ckel  
approximation \cite{Lu}.

\section{Summary and Concluding Remarks}

We summarize our results.\\ 
(i) In the first part, 
we have examined the concentration profile around a 
charged particle in a near-critical polar 
binary mixture,  as can be seen in Figs.1 and 2. 
We have started  with  the simple Ginzburg-Landau 
free energy (2.3)  supplemented with the relations (2.1) and (2.2) of 
electrostatics.  Preferential solvation can occur in the presence 
of a concentration-dependent dielectric constant 
and becomes strong when the parameter $A$ defined by 
(2.22) satisfies  the condition (2.24). 
As the critical point is approached, 
the concentration 
has a long-range Ornstein-Zernike tail (2.30) 
in an intermediate range $a<r<\lambda$ as 
demonstrated in Fig.3. We plot 
the coefficient $B$ vs $A$ 
in Fig.4 and the crossover length 
$\lambda$ vs $A$ in Fig.5. At the critical point 
$\psi(r)$ is long-ranged in the form of (2.43) 
as demonstrated in Fig.6.
The solvation free energy $F_{\rm sol}$ 
behaves as a function of $A$ as in  Fig.7.\\
(ii)  
In the second part of this work we have 
presented the coarse-grained free energy as in (3.4) 
where the order parameter and the ion densities 
are strongly coupled in the bilinear form. 
We have related the coupling constants $w_1$ and $w_2$ 
in terms of $B_1$ and $B_2$, the coefficients of the 
Ornstein-Zernike tails for the ion, 1 and 2, respectively.  
Furthermore, if this coupling is significantly 
different between the two species of the ions, 
a charge-density-wave  phase can be realized.

We then give some remarks.\\
(i) We have neglected the dielectric saturation 
near the ion \cite{Abraham,nonlinear}. If this effect is 
taken into account, the growth of the coefficient $B$ 
with increasing $A$ would be weaker than in Fig.4.\\
(ii) In future experiments, 
the structure factor (3.18) should be 
observed. There could be the case  $\gamma_{\rm p}>1$, 
where the mesoscopic 
phase is formed at low temperatures,   for  pairs of 
small cations like Li$^+$ or Al$^{ 3+}$ and 
relatively large anions \cite{Is}.\\
(iii) Dynamical problems remain largely unsolved. 
For example, we are interested in 
the response of ions against ac electric field 
in near-critical polar binary mixtures. 
In the strong solvation condition 
the ion mobility should strongly depend on 
$\omega\tau_\xi$ where $\omega$ is the 
  frequency of the field and $\tau_\xi$ is  the  
order parameter life time 
($=6\pi\eta\xi^3/k_{\rm B}T$ with $\eta$ being the 
shear viscosity \cite{Onukibook}). 
Furthermore, even if $\omega\tau_\xi \ll 1$, 
the   large-scale solvation cloud on the scale of $\xi$  
should become   nonlinearly dependent on the 
applied field.\\
(iv) Solvation or polarization effects   
due to inhomogeneous dielectric constant 
(or dielectric  tensor for liquid crystals)  
can strongly influence 
the phase transition behavior in
 complex fluids including 
polyelectrolytes, charged gels, 
and liquid crystals containing ions or 
charged colloids \cite{OnukiL}.  The 
importance of this effect 
in  polyelectrolytes  has recently 
been  pointed out by 
Kramarenko {\it et al.} \cite{Kra}\\
(v) A similar Ginzburg-Landau theory can 
easily be developed on  solvation effects in 
polar one-component fluids. 
 For example, 
 we may start with the free energy density 
in the van der Waals theory \cite{Onukibook} 
in place of the Hildebrand free energy density  (2.4).  
As regards statics salient features in one-component 
fluids are nearly  the same as those in 
polar binary mixtures,  
but dynamics of critical 
electrostriction should 
be very different between the two cases \cite{Onukibook}.

\vspace{2mm} 
{\bf Acknowledgments}
\vspace{2mm}

One of the authors (A.O.)  would like to thank  
K. Orzechowski, M. Misawa,  and 
M. Anisimov  for valuable discussions 
 on the ion  effects in near-critical fluids.  
This work is supported by 
Grants in Aid for Scientific 
Research 
and for the 21st Century COE "Center 
for Diversity and Universality in Physics" from the Ministry of Education, 
Culture, Sports, Science and Technology (MEXT) of Japan.

\vspace{2mm} 
{\bf Appendix}\\
\setcounter{equation}{0}
\renewcommand{\theequation}{A.\arabic{equation}}

In equilibrium $\phi(r)$  around a charged particle
 is determined by 
\be 
\hat{f}' -C\nabla^2\phi -
\frac{\ve_1}{8\pi\ve^2 r^4} Z^2e^2\theta^2=0,
\en  
where $\hat{f}$ is  defined 
in (2.16). Multiplying  $\phi'=d\phi/dr$ and 
integrating over $r$ in the range $[r,\infty]$, 
we obtain   
\be 
\hat{f}+ \frac{Z^2e^2}{8\pi\ve r^4} \theta^2
-\frac{C}{2} (\phi')^2 = - \int_r^\infty dr_1H(r_1),
\en 
where 
\be 
H(r)=\frac{2C}{r} (\phi')^2 +  
\frac{Z^2e^2}{8\pi\ve} \frac{d}{dr}\bigg 
(\frac{\theta^2}{r^4}\bigg ) .
\en 
Integration of (A.2) over the whole space 
yields (2.17) with the aid of $\int_0^\infty dr r^2 
\int_r^\infty dr_1 H(r_1)= \int_0^\infty dr r^3 H(r)/3$.  
.


\end{multicols}

\begin{figure}[t]
\vspace{1cm}
\epsfxsize=4in 
\centerline{\epsfbox{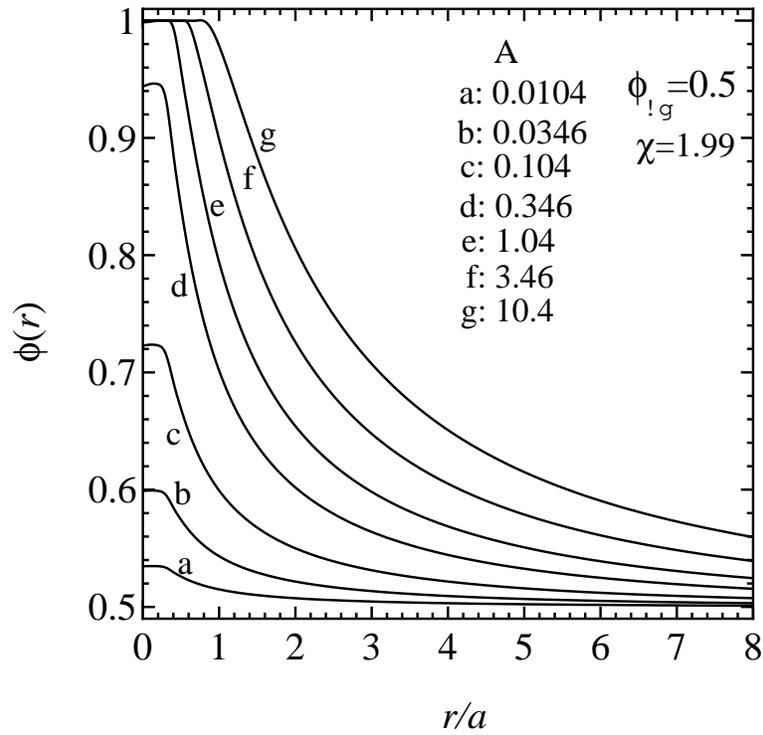}}
\caption{\protect
Concentration profile $\phi(r)$  around an ion 
for various $A$ at $\chi=1.99$ where 
$\xi=10a$ and $\phi\rightarrow \phi_\infty= 1/2$ far from the ion. }
\label{1}
\end{figure}

\begin{figure}[t]
\vspace{1cm}
\epsfxsize=3.4in 
\centerline{\epsfbox{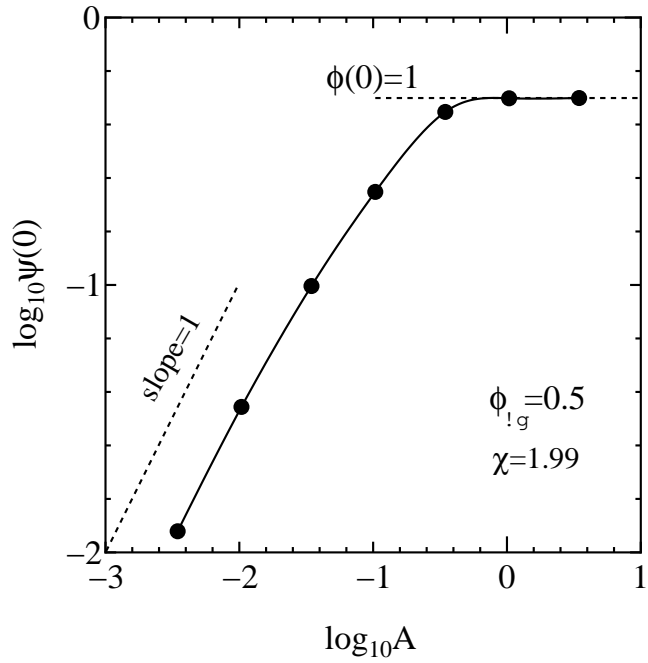}}
\caption{\protect
Concentration  deviation  
$\psi(0)=\phi(0)-1/2$ at the ion center 
for various $A$ at $\chi=1.99$. }
\label{2}
\end{figure}

\begin{figure}[t]
\vspace{1cm}
\epsfxsize=4in 
\centerline{\epsfbox{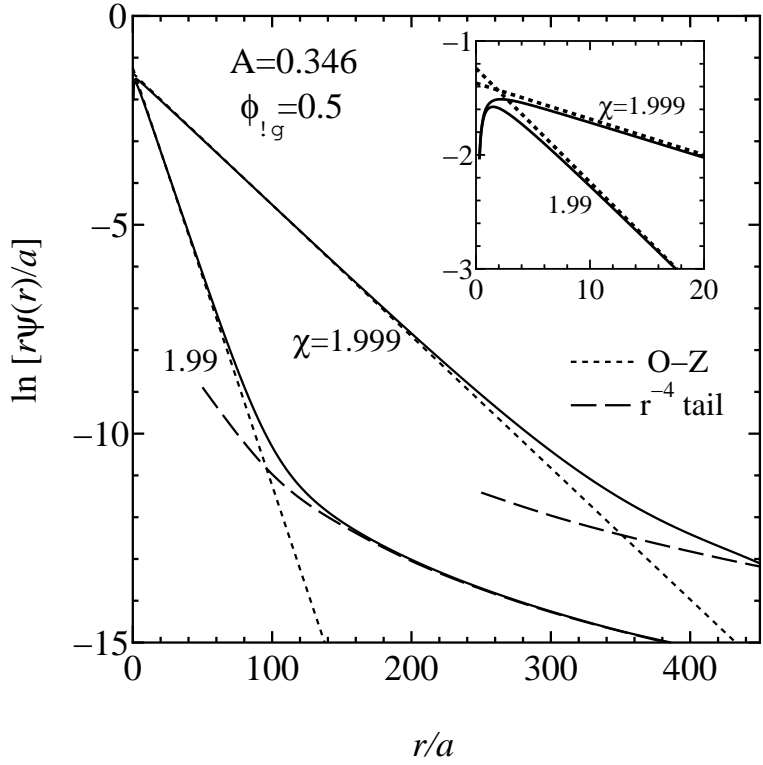}}
\caption{\protect
Numerical $r\psi(r)/a$ vs $r/a$ 
on a semi-logarithmic scale at $A=0.346$ 
for $\chi=1.99$ (lower 
solid line) and $1.999$ (upper solid  line) 
with $\phi_\infty=1/2$. 
The numerical $\psi(r)=\phi(r)-1/2$ can  be 
 excellently fitted to the 
Ornstein-Zernike form (corresponding to dotted lines) in 
the intermediate range  $a \ls r\ls \lambda$,  while 
it can be fitted to the $r^{-4}$ tail (2.37) (corresponding 
to broken lines) for $r \gs \lambda$. The curve of $\lambda$ vs $A$ will 
be given in Fig.5.  
 }
\label{3}
\end{figure}

\begin{figure}[t]
\vspace{1cm}
\epsfxsize=3.4in 
\centerline{\epsfbox{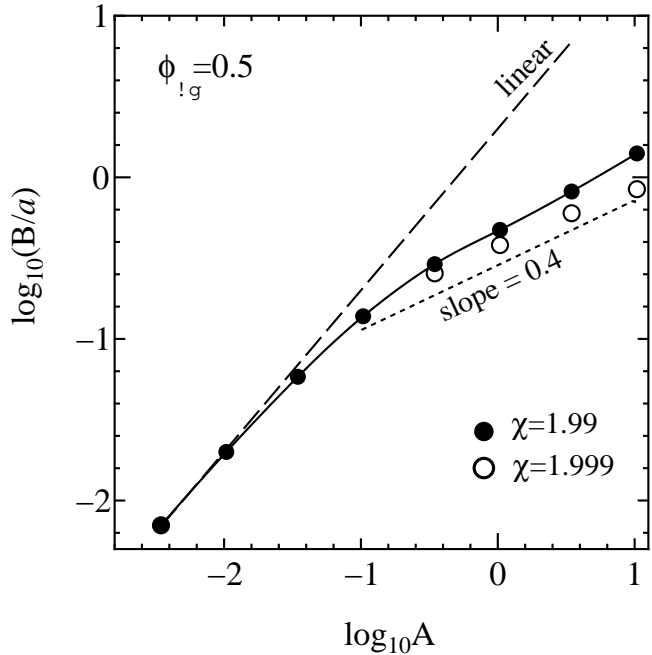}}
\caption{\protect
Coefficient $B$ of the Ornstein-Zernike tail 
 in the concentration profile as a function of $A$ 
at $\phi_\infty=0.5$ for $\chi=1.99$ and $1.999$. 
It exhibits only weak dependence on $2-\chi\propto T-T_{\rm c}$. }
\label{4}
\end{figure}

\begin{figure}[t]
\vspace{1cm}
\epsfxsize=3.4in 
\centerline{\epsfbox{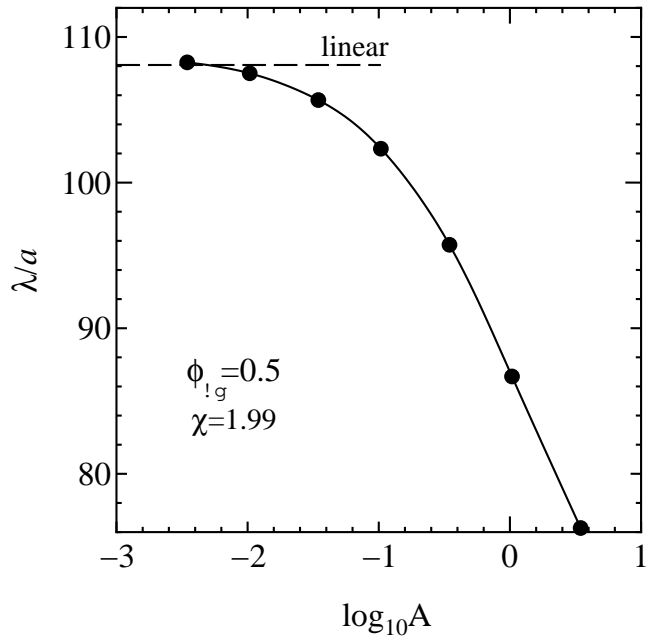}}
\caption{\protect
Crossover length 
$\lambda$ from the Ornstein-Zernike 
tail  (2.30) to the $r^{-4}$ tail (2.37) 
at $\phi_\infty=0.5$ for $\chi=1.99$ where 
$\xi=10a$. }
\label{5}
\end{figure}

\begin{figure}[t]
\vspace{1cm}
\epsfxsize=4in 
\centerline{\epsfbox{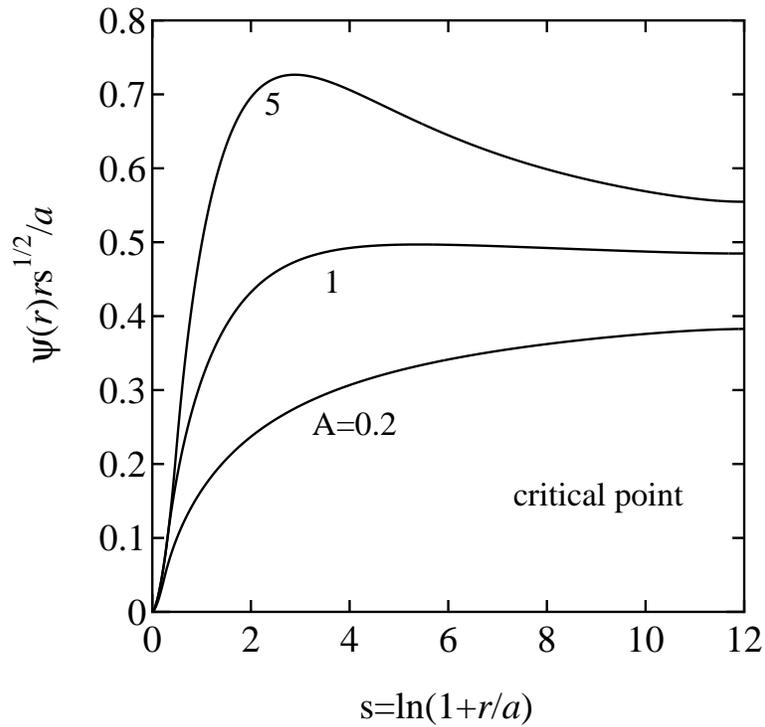}}
\caption{\protect
$\psi(r)rs^{1/2}$ vs $s=\ln(1+r/a)$ 
at the critical point   
for  $A=0.2,1$, and $5$. The curves appear to approach  
$\sqrt{3}/4$ 
very slowly 
supporting (2.43).} 
\label{6}
\end{figure}

\begin{figure}[t]
\vspace{1cm}
\epsfxsize=4in 
\centerline{\epsfbox{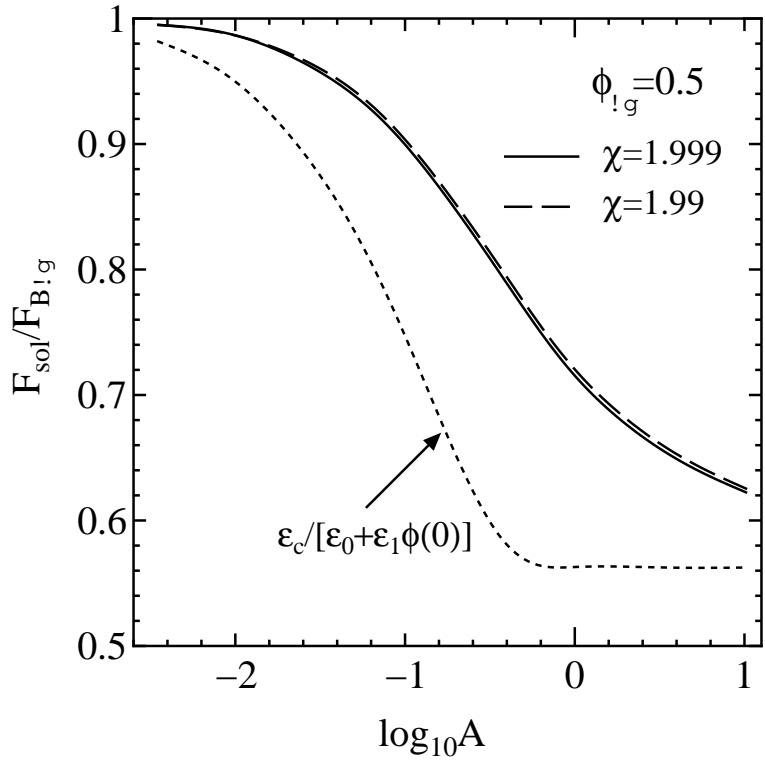}}
\caption{\protect
Solvation free energy $F_{\rm sol}$ in 
(2.17) divided by $F_{{\rm B}\infty}$ in (2.13) 
as a function of $A$. The two curves for $\chi=1.99$ 
(broken  line) and $1.999$ (solid line) 
are almost identical. 
The dotted line represents $F_{{\rm B}0}/F_{{\rm B}\infty}$ 
at $\chi=1.99$ (see (2.16)).} 
\label{6}
\end{figure}

\begin{figure}[t]
\vspace{1cm}
\epsfxsize=3.4in 
\centerline{\epsfbox{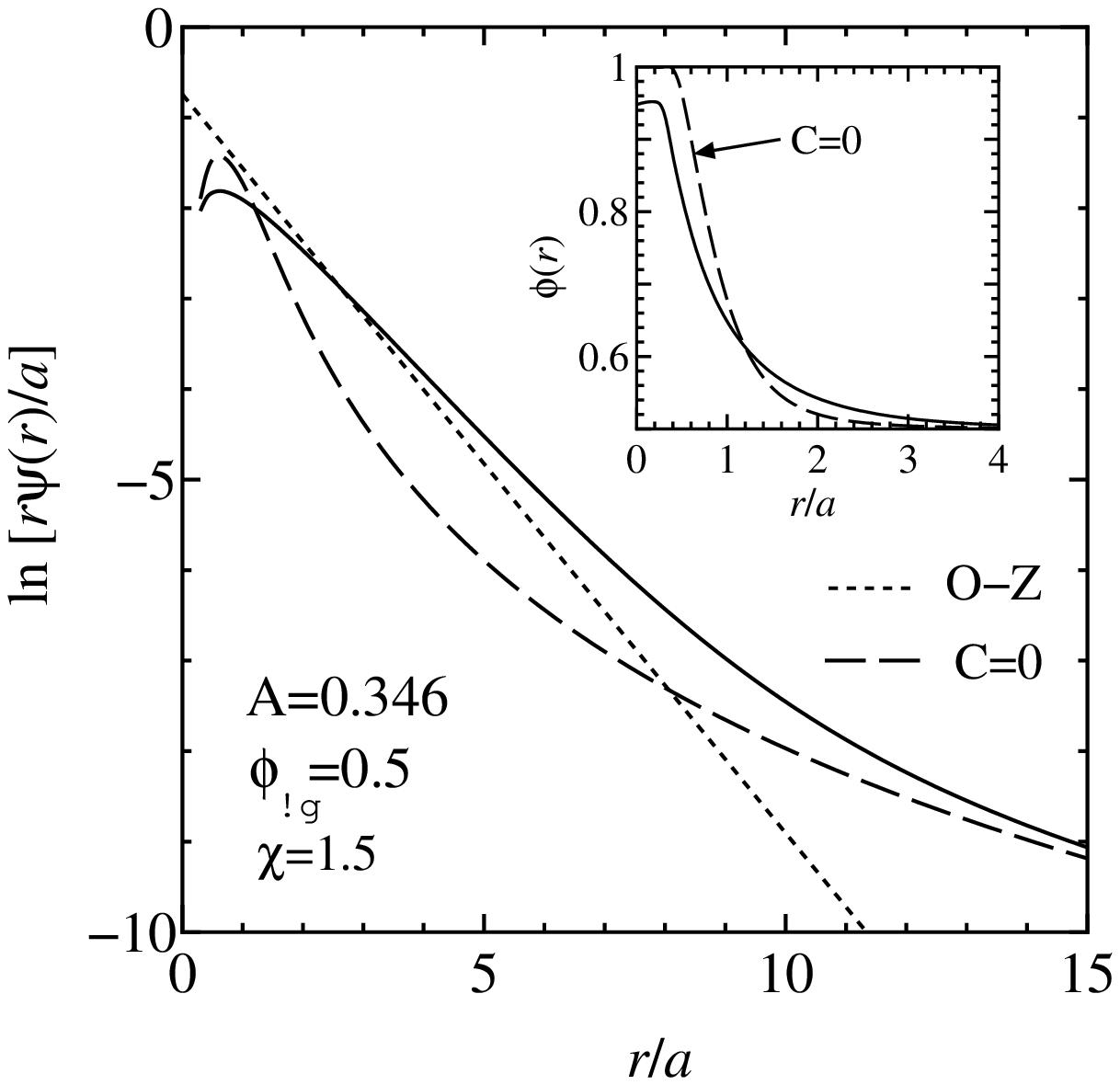}}
\caption{\protect
$r\psi(r)$ vs $r/a$ on a semi-logarithmic scale 
 around an ion far above the critical point 
with  $\chi=1.5$ and $\phi_\infty=0.5$.
The profile can be fitted neither  to the Ornstein-Zernike 
form (dotted line) nor to the local equilibrium curve 
without the gradient  term (broken line).  The inset 
displays $\phi(r)$ vs $r/a$ near the ion 
on a linear scale.  } 
\label{8}
\end{figure}


\begin{references}
\bibitem{commentG} 
If we set 
$C=2\alpha k_{\rm B}T/a$, our results near the critical 
point are those at $\alpha=1$. 
Obviously, results  for general $\alpha$ 
can be obtained from ours by simple scale 
changes. For example, Fig.4 becomes 
the relation of  $B\alpha^{1/2}/a$ vs 
$A\alpha^2$.
\end{references}
\end{document}